# Practical applications of Set Shaping Theory in Huffman coding

C.Schmidt, A.Vdberg, A. Petit

Abstract: One of the biggest criticisms of the Set Shaping Theory is the lack of a practical application. This is due to the difficulty of its application. In fact, to apply this technique from an experimental point of view we must use a table that defines the correspondences between two sets. However, this approach is not usable in practice, because the table has $|A|^N$ elements, with $|A|$ number of symbols and N length of the message to be encoded. Consequently, these tables can be implemented in a program only when $|A|$ and N have a low value. Unfortunately, in these cases, there are no compression algorithms with such efficiency as to detect the improvement introduced by this method. In this article, we use a function capable of performing the transform without using the correspondence table; this allows us to apply this theory to a wide range of values of $|A|$ and N. The results obtained confirm the theoretical predictions.

## Introduction

Set Shaping Theory studies the bijection functions *f* that transform a set $X^N$ of strings of length *N* into a set $Y^{N+K}$ of strings of length *N+K* with K and $N \in \mathbb{N}^+, |X^N| = |Y^{N+k}|$ and $Y^{N+K} \subset X^{N+K}$. The set $X^N$ contains all the sequences of length N that can be generated from an alphabet A, so $|X^N| = |A|^N$. The simplest practical application of this function is to use a table that defines the correspondences between the two sets. However this approach is not usable in practice, because the table has $|A|^N$ elements therefore, these tables can be implemented in a program only when $|A|$ and N have a low value. Unfortunately, in these cases, there are no compression algorithms with such efficiency as to detect the improvement introduced by the transform. In this article, we use a function capable of performing the transform without using the correspondence table; this allows us to apply this theory to a wide range of values of $|A|$ and N. The results obtained confirm the theoretical predictions, this experimental confirmation is fundamental because one of the major criticalities of this method was precisely the lack of a practical application.

# Brief description of the Set Shaping Theory

The Set Shaping Theory [1] has as its objective the study and application in information theory of bijection functions $f$ that transform a set $X^N$ of strings of length $N$ into a set $Y^{N+K}$ of strings of length $N+K$ with $K$ and $N \in \mathbb{N}^+$, $|X^N| = |Y^{N+k}|$ and $Y^{N+K} \subset X^{N+K}$.

The function $f$ defines from the set $X^{N+K}$ a subset of size equal to $|X^N|$. This operation is called "Shaping of the source", because what is done is to make null the probability of generating some sequences belonging to the set $X^{N+K}$.

The parameter $K$ is called the shaping order of the source and represents the difference in length between the sequences belonging to $X^N$ and the transformed sequences belonging to $Y^{N+k}$.

Definition: Given a sequence x, we call Cs(x) the coded sequence in which the symbols are replaced by a uniquely decodable code [2].

Codes of this type have the characteristic that no codeword is the prefix of another codeword. Thus, a code is uniquely decodable if it is possible to decode each transmitted character unambiguously, without ambiguity. These codes are very important because many theorems such as Shannon's first theorem [3] refer to this type of codes.

Definition: given a sequence x of random variables of length N, we call the coding limit Lc(x) the function defined as follows:

$$Lc(x) = -\sum_{i=1}^{N} \log_2 p(x_i)$$

With $p(x_i)$ we mean the "actual frequency" of the symbol $x_i$ in the sequence.

The function Lc tells us that a sequence x of random variables of length N on average cannot be encoded in less than Lc(x) bits. At most, there may be a function that can transform the sequence x into a new sequence that reduces the value of Lc 50% of the time and increases it the remaining 50%, obtaining no gain on average.

In Set Shaping Theory this function is referred to as the sequence information content. In this article, we have chosen not to use this definition because it can be risky to talk about information without knowing the source that generated the message. This comment does not mean that the definition given in the Set Shaping Theory is incorrect; simply we have chosen a less ambiguous definition linked to an experimental limit.

Given a set $X^N$ which contains all the sequences of length N that can be generated, therefore with dimension $|X^N| = |A|^N$. The Set Shaping Theory [4] tells us that when $|A| > 2$ exists a set $Y^{N+K}$ of dimension $|A|^N$ consisting of sequences having alphabet A and length N+K, in which the average value of Lc(y) is less than the average value of Lc(x) calculated on the sequences belonging to $X^N$.

Since the two sets have the same dimension, it is possible to put the sequences belonging to the two sets into a one-to-one relationship. Consequently, this function would allow us to exceed the limit defined by the parameter Lc(x) obtaining a result of enormous interest. In practice, by applying this function it is possible with a probability greater than 50% to transform a sequence x belonging to $X^N$ into a new sequence f(x)=y belonging to $Y^{N+K}$ which can be encoded with a smaller number of bits than to Lc(x).

**Description of the program and the results obtained**

Appendix A shows the Matlab code program that can be downloaded from the following link:

https://www.mathworks.com/matlabcentral/fileexchange/116025-test-new-data-compression-technique

The program performs the following experiment:

Generate a random sequence with uniform distribution (symbol emission probability 1/|A|) with alphabet A and length N.

For example, we generated the following sequence x with N=10 and |A|=5.

2223344551

We calculate the frequencies of the symbols in the sequence.

Symbol 1 frequency 1/10

Symbol 2 frequency 3/10

Symbol 3 frequency 2/10

Symbol 4 frequency 2/10

Symbol 5 frequency 2/10

Calculate the parameter Lc(x), as defined previously.

$$Lc(x) = \log_2 \frac{3}{10} - \log_2 \frac{3}{10} - \log_2 \frac{3}{10} - \log_2 \frac{2}{10} - \log_2 \frac{2}{10} - \log_2 \frac{2}{10} - \log_2 \frac{2}{10}$$
$$- \log_2 \frac{2}{10} - \log_2 \frac{2}{10} - \log_2 \frac{1}{10} = 22.5 \; bit$$

This value defines the length limit of the encoded sequence in which symbols are substituted for codewords (Uniquely Decodable code).

We perform a transform on sequence x generating a new sequence f(x). Let us take, as an example, a function that transforms the sequence x into the following sequence f(x) with N=11.

11111122435

We calculate the frequencies of the symbols in the sequence.

Symbol 1 frequency 6/11

Symbol 2 frequency 2/11

Symbol 3 frequency 1/11

Symbol 4 frequency 1/11

Symbol 5 frequency 1/11

We encode the sequence f(x) using the Hufmman encoding [5] in which the codewords are optimized for the frequencies that we have calculated (symbol 1 6/11, symbol 2 2/11, symbol 3,4,5 1/11). The codewords found using these frequencies are:

Symbol 1 codeword 1

Symbol 2 codeword 000

Symbol 3 codeword 001

Symbol 4 codeword 010

Symbol 5 codeword 011

The coded sequence becomes:

111111000000010001011

A sequence of 21 bits length, therefore, a value less than the 22.5 bits that constituted the encoding limit Lc(x) of the initial sequence.

The program performs these steps a number of volts indicated by the parameter history. Each time that the transformed sequence f(x) is encoded with a number of bits less than Lc(x) the value of the counter cs increases by one unit. When this cycle ends, we calculate the probability that the transformed sequence f(x) can be encoded Cs(f(x)) with a number of bits less than Lc(x).

The table shows the results for some settings. The first column reports the parameter ns which indicates the number of symbols of the random sequences with uniform distribution generated, therefore represents $|A|$. The second column reports the length N of the generated messages and the third column the probability ps that the encoding of the transformed message Cs(f(x)) has a length less than Lc(x).

| ns  | N    | Pr  |
|-----|------|-----|
| 40  | 80   | 69% |
| 50  | 100  | 72% |
| 60  | 120  | 80% |
| 500 | 1000 | 88% |

*Table 1: Results obtained for different settings of the parameters ns and N.*

The results confirm the theoretical predictions, In fact, the transformed sequences f(x) can be encoded with a number of bits lower than the Lc(x) of the initial sequence x with a probability greater than 50%. For example, with $|A|$=40 and N=80 there is a probability of about 79% that the transformed sequence can be encoded with a bit number lower than the Lc(x) value of the initial sequence x.

## Conclusion

This article reports the first practical application of set shaping theory. Being able to carry out this method from a practical point of view means being able to develop a function f that transforms a set $X^N$ of strings of length N into a set $Y^{N+K}$ of the same size of strings of length N+K which minimize the function Lc(x). The development of a function of this type represents a difficult problem to solve and this is the reason for the absence of an experimental application of set shaping theory. Consequently, this lack also determines the greater criticality of this theory.

The solution of this problem allowed us to verify experimentally the theoretical predictions. In fact, the results obtained confirm the prediction made by the set shaping theory relating to the possibility of exceeding the coding limit Lc(x) in the conditions defined by the experiment.

We do not express ourselves on the importance of the result obtained, because the purpose of this article is only the experimental application of the set shaping theory and the sharing of the code used to obtain the results presented. By sharing the function that performs the transform, we hope to help anyone who wants to study this new theory.

## Reference

1. Solomon Kozlov. Introduction to Set Shaping Theory. *ArXiv, abs/2111.08369, 2021*.

2. Ranjan Bose, Information theory coding and cryptography, Boston : McGraw-Hill, 2003.

3. Shannon, C. E., & Weaver, W. (1949). *The mathematical theory of communication*. Urbana: University of Illinois Press.

4. Solomon Kozlov. Use of Set Shaping theory in the development of locally testable codes. arXiv:2202.13152. February 2022.

5. Huffman, D. (1952). A Method for the Construction of Minimum-Redundancy Codes. Proceedings of the I.R.E. 40 (9):1098 1101. doi: 10.1109/ JRPROC.1952.273898.

Contact info

C.Schmidt: Christian.Schmidt55u@gmail.com

A.Vdberg: adrain.vdberg66@yahoo.com

A. Petit: alix.petitaus@gmail.com

# Appendix A

The Matlab program and the functions that perform the transform of the sequence x and the inverse transform can be downloaded from the following link on Matlab fileexchange;

https://www.mathworks.com/matlabcentral/fileexchange/116025-test-new-data-compression-technique

```
%                f_Statistics_Set_Shaping_Theory_Huffman_l
%                          Contact info
%                   Christian.Schmidt55u@gmail.com
%                       adrain.vdberg66@yahoo.com
%
%%%%%%%%%%%%%%%%%%%%%%%%%%%%%%%%%%%%%%%%%%%%%%%%%%%%%%%%%%%%%%%%%%%%%%%%%%
%                           Definition
%
% Definition: Given a sequence x we call Cs(x) the coded sequence in which
% the symbols are replaced by a uniquely decodable code.
%
% Definition: given a sequence x of random variables of length N, we call
% the coding limit Lc(x) the function defined by the following example:
%
% given a sequence:
% 123445
%
% We calculate the frequencies of the symbols present in the sequence.
%
%  symbol 1  frequencies 1/6
%  symbol 2  frequencies 1/6
%  symbol 3  frequencies 1/6
%  symbol 4  frequencies 2/6
%  symbol 5  frequencies 1/6
%
%  Lc(x)=-log2(1/6)-log2(1/6)-log2(1/6)-log2(2/6)-log2(2/6)-log2(1/6)=13.5 bit
%
% 13.5 bit represents the minimum length of the coded sequence
% in which the symbols have been replaced with uniquely decodable code
%%%%%%%%%%%%%%%%%%%%%%%%%%%%%%%%%%%%%%%%%%%%%%%%%%%%%%%%%%%%%%%%%%%%%%%%%%
% The program performs the following operations:
% 1) generates a random sequence x with uniform distribution
% 2) calculate the frequencies of the symbols present in the sequence x
% 3) use this information to calculate the coding limit Lc(x)
% 4) apply the transform f(x)
% 5) code the transformed sequence
% 6) compares the coding limit Lc(x) of the generated sequence
%    with the length of the encoded transformated sequence cs(f(x))
% 7) repeats all these steps a number of times defined by the parameter history
% 8) display the average values obtained
%%%%%%%%%%%%%%%%%%%%%%%%%%%%%%%%%%%%%%%%%%%%%%%%%%%%%%%%%%%%%%%%%%%%%%%%%%
%                           Important
%
% If you change the length of the sequence and the number of the generated
% symbols, you have to be careful that the Huffman encoding approximates the
% coding limit of the sequence by about one bit. if you take too
% long sequences the Huffman algorithm becomes very inefficient therefore,
```

```matlab
% it cannot detect the advantage obtained by the transformation.
% As a general rule, if you take ns symbols the length of the sequence
% must be about 2*ns, in this case the Huffman encoding approximates
% the coding limit of the sequence by about one bit.
%%%%%%%%%%%%%%%%%%%%%%%%%%%%%%%%%%%%%%%%%%%%%%%%%%%%%%%%%%%%%%%%%%%%%%
%                Results for different settings
%
%  The fSST2 and invfSST2 function only works when the number of symbols ns
%  is greater or equal than 20 and less or equal than 500 and the length of
%  the sequence N is greater or equal than 40 and less or equal than 1000.
%  So, it is recommended to generate random sequences with a number of
%  symbols and length greater than Ns=20 and N=40.
%
%  Ps=probability with which the transformed sequence f(x) can be encoded
%  using a uniquely decodable code (Huffman coding) with lower bit number
%  than coding limit Lc(x) of the initial sequence x.
%
%  ns= number of symbols
%  N=length of the sequence
%
%  ns    N      Ps
%  40    80     69%
%  50    100    72%
%  60    120    80%
% 500   1000    88%
%
%  As you can see, increasing the symbol number increases the probability Ps
%%%%%%%%%%%%%%%%%%%%%%%%%%%%%%%%%%%%%%%%%%%%%%%%%%%%%%%%%%%%%%%%%%%%%%%
clear all;
history=1000;
ns=40;
len=80;
cs=0;
totcodel=0;
totlc=0;
tottlc=0;
itnent=0;
lc=0;
tlc=0;
itlc=0;

for i=1:history

 % Generation of the sequence with a uniform distribution

 symbols=1:ns;
 prob(1,1:ns)=1/ns;
 seq=randsrc(1,len,[symbols; prob]);

 % coding limit Lc(x)

 lc=0;

 for i2=1:len

  sy=seq(1,i2);
  fs=nnz(seq==sy)/len;
  lc=lc-log2(fs);

 end
```

```matlab
% Start trasformation

mcodel=10000;

nseq=fSSTt(seq);

% The new sequence is long nlen=len+1

nlen=len+1;

% coding limit of the transformed sequence of length nlen

 tlc=0;

 for i2=1:nlen

  sy=nseq(1,i2);
  fs=nnz(nseq==sy)/nlen;
  tlc=tlc-log2(fs);

 end

% Having transformed the sequence, we have to redefine the length of the
% vectors that are used in the encoding

index=0;

for i2=1:ns

 fs=nnz(nseq==i2)/nlen;

 if fs > 0

   index=index+1;

 end

end

c=zeros(index,1);
vs=zeros(index,1);
index=0;

% We calculate the frequencies of the symbols in the transformed sequence

for i2=1:ns

 fs=nnz(nseq==i2)/nlen;

 if fs > 0

  index=index+1;
  c(index)=nnz(nseq==i2)/nlen;
  vs(index)=i2;

 end

end

% We code the transformed sequence
```

```matlab
  counts=c;

  tdict=huffmandict(vs,counts);
  tcode=huffmanenco(nseq,tdict);

  bcode=de2bi(tcode);
  tcodel=numel(bcode);

  % If the length of the encoded message of the transformed sequence is less
  % than coding limit Lc(x) of the original sequence x, we increase the counter
  % cs by one

  if tcodel < lc

    cs=cs+1;

  end

  % We apply the inverse transform and we obtain the initial sequence

  iseq=invfSSTt(nseq);

  % We check that the obtained sequence is equal to the initial sequence

  flag=isequal(seq,iseq);

  if flag == false

     fprintf('Error, sequence not equal to the initial sequence\n');

  end

  totcodel=totcodel+tcodel;
  totlc=totlc+lc;
  tottlc=tottlc+tlc;

end

% We calculate the average of the coding limit Lc(x) of the generated sequences
% x,the average of the coding limit Lc(f(x)) of the transformed sequences f(x)
% and the average of the length of the encoded transformed sequence Cs(f(x))

 medlc=totlc/history;
 medcodel=totcodel/history;
 medtlc=tottlc/history;

% We calculate the percentage of sequences where the length of the encoded
% transformed sequence Cs(f(x)) is less than the coding limit Lc(x) of the
% generated sequence x

 pcs=(cs/history)*100;

% We display the average values obtained

 fprintf('The average of the coding limit Lc(x) of the generated sequences x\n');
 medlc

 fprintf('The average of the length of the encoded transformed sequence Cs(f(x)) \n');
 medcodel
```

```
 fprintf('The average of the coding limit Lc(f(x)) of the transformed sequences 
f(x)\n');
 medtlc

 fprintf('Number of sequences where the length of the encoded transformed sequence 
Cs(f(x)) is less than the coding limit Lc(x) of the generated sequence x\n');
 cs

 fprintf('There is a percentage of %2.0f%% that length of the encoded transformed 
sequence Cs(f(x)) is less than the coding limit Lc(x) of the generated sequence 
x\n',pcs);
```